\documentclass[prl,twocolumn,superscriptaddress,nofootinbib, ]{revtex4-1}

\usepackage{graphicx}
\usepackage{amssymb}
\usepackage{amsmath}

\allowdisplaybreaks[4]

\def\beq{\begin{equation}}
\def\eeq{\end{equation}}
\def\bea{\begin{eqnarray}}
\def\eea{\end{eqnarray}}
\def\met{\not{\!{\rm E}}_T}
\def\metvec{\not{\!\vec{\bf E}}_T}
\def\nn{\nonumber}

\begin{document}

\title{Testing LHT at the LHC Run-II}

\author{Qing-Hong Cao}
\email{qinghongcao@pku.edu.cn}
\affiliation{Department of Physics and State Key Laboratory 
of Nuclear Physics and Technology, Peking University, Beijing 100871, China}
\affiliation{Collaborative Innovation Center of Quantum Matter, Beijing 100871, China}
\affiliation{Center for High Energy Physics, Peking University, Beijing 100871, China}

\author{Chuan-Ren Chen}
\email{crchen@ntnu.edu.tw}
\affiliation{Department of Physics, National Taiwan Normal University, Taipei 116, Taiwan}

\author{Yandong Liu}
\email{ydliu@pku.edu.cn}
\affiliation{Department of Physics and State Key Laboratory of Nuclear Physics
and Technology, Peking University, Beijing 100871, China}

\begin{abstract}
We study the Littlest Higgs model with T-parity (LHT) in the process of $pp \to W_H^+W_H^- \to W^+W^- A_H A_H$ at the 14 TeV LHC. With the $W$-jet tagging technique, we demonstrate that the bulk of the model parameter space can be probed at the level of more than $5\sigma$ in the signature of two fat $W$-jets plus large missing energy. Furthermore, we propose a novel strategy of measuring the principle parameter $f$ that is crucial to testify the LHT model and to fix mass spectrum, including dark matter particle. Our proposal can be easily incorporated into current experimental program of diboson searches at the LHC Run-II. 
\end{abstract}

\maketitle

\noindent{\bf Motivation:~}%
The mystery of why the mass of Higgs boson is at the weak scale remains after the Higgs discovery at the Large Hadron Collider (LHC). One possibility is given in the framework of Little Higgs models in which the Higgs boson emerges as a pseudo Nambu-Goldstone boson in the mechanism of collective symmetry breaking~\cite{ArkaniHamed:2001nc,ArkaniHamed:2002qy}.  
The original construction of Little Higgs model suffers severely from electroweak precision tests that demand the collective symmetry breaking scale $f(=\Lambda/4\pi)$ to be large~\cite{Csaki:2002qg}. Therefore, the model reintroduces the fine-tunning and has little relevance to the current high energy collider physics program. The stringent constants can be naturally released when a discrete symmetry, called T-parity, is imposed~\cite{Cheng:2003ju,Cheng:2004yc,Low:2004xc}. All the corrections to electroweak observables are loop-induced. The value of $f$, then, can be as low as $500~{\rm GeV}$~\cite{Hubisz:2005tx}, and the masses of new heavy resonances are below TeV. 

In this Letter we consider the ``Littlest'' Higgs model with T-parity (LHT), which is based on an $SU(5)/SO(5)$ nonlinear sigma model whose low energy Lagrangian is described in details in Ref.~\cite{Hubisz:2004ft,Belyaev:2006jh,Reuter:2013iya}. In the model under the T-parity transformation the SM particles are neutral while all the new particles are odd, except a top quark partner that cancels out the SM top quark's contribution to the quadratic divergence in radiative corrections of mass parameter of the Higgs boson. The characteristics of the LHT are the dependence of few free parameters and the tight mass relation between heavy gauge bosons. For example, after the electroweak symmetry breaking, the masses of the T-parity partners of the photon $(A_{H})$ and $W$-boson $(W_{H})$ are generated as 
\bea
M_{A_{H}}=\frac{g^\prime f}{\sqrt{5}}\left(1-\frac{5v^{2}}{8f^{2}}\right),~
M_{W_{H}}=gf\left(1-\frac{v^{2}}{8f^{2}}\right),
\eea
where $v$ is the vacuum expectation value, and $g$ and $g^\prime$ are the gauge couplings of $SU(2)_L$ and $U(1)_Y$, respectively. Because of the smallness of the $U(1)_{Y}$ gauge coupling constant $g'$, the T-parity partner of the photon $A_{H}$ tends to be the lightest T-odd particle (LTP) in the LHT, which serves the dark matter (DM) candidate~\cite{Asano:2006nr}.
Given the $SU(2)_{L}$ gauge coupling constant $g$ and the vacuum expectation value $v$ ($\simeq 246$ GeV) being measured in SM electroweak processes, the measurement of $M_{W_H}$ could determine the value of $f$, the most important parameter of the LHT model. That in turn determines $M_{A_H}$ (the DM mass), which is crucial for other experiments of dark matter searches, e.g. direct and indirect searches for dark matter. We demonstrate that the scale $f$ can be determined through the production of charged heavy gauge boson pair ($W_H^+ W_H^-$) at the LHC.

\begin{figure}[b]
\includegraphics[scale=0.5]{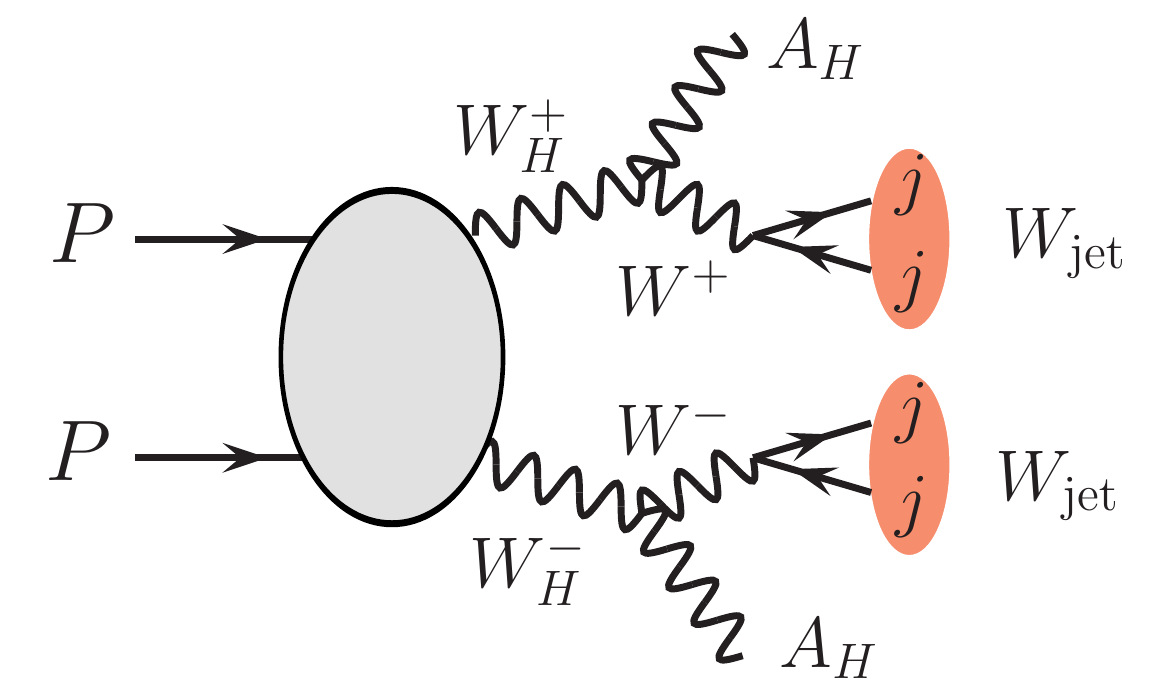}
\caption{Pictorial illustration of the pair production and decay of $W_H$ at the LHC. 
\label{fig:feyn}}
\end{figure}

The $W_H$ boson almost entirely decays into a pair of $W$ and $A_H$ bosons in the LHT model~\cite{Cao:2007pv}. Therefore, the collider signature of the $W_H^+W_H^-$ production is controlled by decay products of the $W$ bosons from the $W_H$ decays. Both the leptonic and hadronic decay of the $W$ bosons in the $W_H^+W_H^-$ production have been studied in Refs.~\cite{Cao:2007pv,SongMing:2012gb}, which pointed out that, owing to invisible DM particles in the final state,  the event reconstruction is difficult. For example, one immediately confronts two difficulties  that preclude the event reconstruction in the hadronic mode: 
i) unknown DM mass and undetectable DM momentum; 
ii) the $W$-boson being highly boosted such that its decay product tend to be highly collimated and hard to be isolated. 
We propose a novel method to overcome those difficulties to measure $M_{W_H}$, $M_{A_H}$ and $f$ in the process of $pp\to W_H^+ W_H^- \to W^+(\to j j)W^-(\to jj) A_H A_H $; see Fig.~\ref{fig:feyn}. 

~\\
\noindent {\bf The Method:~}%
We begin with the difficulty in the $W$-boson reconstruction. In the allowed parameter space of $f$ the $W_H$ boson is much heavier than the $W$  and $A_H$ bosons. Thus, the $W$ boson from the $W_H$ decay is highly boosted such that the two jets of the $W$ boson decay tend to be collimated and form one fat jet in the detector. We name it as a $W$-jet ($W_J$). The substructure of the $W$-jet provides a powerful tool to detect the signal event and suppress background significantly; for example, when the transverse momentum ($p_T$) of the $W$ boson is larger than 200 GeV, the boosted jet algorithm, used to identify $W$ jets, becomes more efficient than the reconstruction of a $W$ boson from two isolated jets~\cite{Khachatryan:2014vla}. The $W$-jet technique has been successfully used in searching for new resonances in the diboson channel by the ATLAS~\cite{Aad:2015owa} and CMS~\cite{Khachatryan:2014hpa,CMS-PAS-EXO-15-002} collaborations. 

Next we deal with the unknown mass of the invisible DM particle. 
The general challenge of determining the mass of new heavy particles that decay into DM is that no mass resonance can be reconstructed because of the lack of kinematic information of DM candidates. In the SM processes with two or less neutrinos in the final state, one might be able to use on-shell conditions of those intermediate particles to reconstruct the momenta of the missing neutrinos~\cite{Berger:2010fy,Zhang:2010kr}; for example, the momentum of the neutrino in the top-quark decay, $t\to bW^+(\to \ell^+\nu)$, can be solved from the on-shell condition of the $W$ boson~\cite{Cao:2005pq}. Unfortunately, the on-shell technique does not apply here because $M_{A_H}$ is unknown. With two DM particles and no neutrino in the final state, the so-called $M_{T2}$ variable is designed in such a way that its distribution will be bound from above by the true mass of the mother particle~\cite{Lester:1999tx}. 
The definition is given as
\begin{widetext}
\beq
m_{T2}(p_{\rm vis}^{(1)}, p_{\rm vis}^{(2)},\metvec; m_{inv})
\equiv \min_{\metvec={\not{\vec{\bf E}}}_{T}^{(1)}+{\not{\vec{\bf E}}}_{T}^{(2)}}
\Bigl[ \max\left\{ 
m_T(p^{(1)}_{\rm vis}, \not{\!\vec{\bf E}}_{T}^{(1)};m_{\rm inv} ), 
m_T(p^{(2)}_{\rm vis}, \not{\!\vec{\bf E}}_{T}^{(2)};m_{\rm inv} )  
\Bigr\}\right], 
\label{eq1}
\eeq
\end{widetext}
where
\bea
&&m_T(p_{\rm vis}^i, \not{\!\vec{\bf E}}_{T}^{i}; m_{\rm inv}) \nn\\
&\equiv&\sqrt{m_{\rm inv}^2 + m_{\rm vis}^2 + 2 (E_T^{\rm vis} \not{\!\rm E}_T^{(i)}-\vec{p}_{\rm vis}^i\cdot \not{\!\vec{\bf E}}_{T}^{i})},
\eea
is the transverse mass of the visible cluster $p_{\rm vis}^i$ and the missing transverse momentum $\not{\!\!\!\vec{\bf E}}_{T}^{i}$. Here, $m_{\rm vis}$ denotes the invariant mass of visible cluster and $m_{\rm inv}$ is the mass of missing particle. In our case, the visible cluster in Eq.~(\ref{eq1}) refers to the $W$-jet and the endpoint of $M_{T2}$ distributions in the upper end determines $M_{W_H}$ when the true mass of the missing dark matter $M_{A_H}$ is inputted. The $M_{T2}$ method is theoretically ideal, however, because there is a lack of mass information of dark matter particle, its application of  determining the mass of mother particle is still limited.

Thanks to the strong correlation between masses of heavy gauge bosons in the LHT model,
\beq
M_{W_H} = (\sqrt{5}g/g^\prime )M_{A_H} \approx 4.2 M_{A_H},
\label{eq:mwhah}
\eeq
we introduce a new variable $M_{T2}^{0}$ to determine the value of $f$ in the $W_H^+W_H^-$ production without knowing $M_{A_H}$.  The variable is defined as
\beq
M_{T2}^{0}\equiv m_{T2}(p_{\rm vis}^{(1)}, p_{\rm vis}^{(2)},\metvec; m_{inv}=0),
\eeq
which, in the limit of $M_{W_H}\gg M_{A_H,W}$, exhibits a distribution with an upper edge at~\cite{Cho:2007dh}
\bea
M_{T2}^{\text{End0}} & \simeq & M_{W_H}-\frac{M_{A_H}^2}{M_{W_H}}\left[1+\frac{M_W^2 }{M_{W_H}^2}+\mathcal{O}\left(\frac{M_{W,A_H}^4}{M_{W_H}^4}\right)\right] 
\nn\\
&\approx & M_{W_H}-\frac{M_{A_H}^2}{M_{W_H}}= \frac{5g^2-g^{\prime2}}{5g}f \left(1-\frac{9}{8}\frac{v^2}{f^2}\right)\nn\\
&\approx & \frac{5g^2-g^{\prime2}}{5g}f \sim 0.62f~.
\label{eq:mt2lht0}
\eea
The measurement of $M_{T2}^{\rm End0}$ determines  the value of $f$, from which the masses of $W_H$ and $A_H$ bosons can be derived.
One then can substitute the derived $M_{A_H}$ into the $M_{T2}$ variable given in Eq.~(\ref{eq1}) to plot the $M_{T2}$ distribution with the upper ending point being the mass of $W_H$. In this way a consistency check of the mass relation given in Eq.~(\ref{eq:mwhah}) can be made, which is of great importance to verify the LHT model. 
Moreover, one can examine the leptonic mode, $W_H W_H \to W^+(\to \ell^+ \nu)W^{-}(\to \ell^{\prime -}\bar{\nu})A_HA_H\to \ell^+ \ell^{\prime -} + \met $, to cross-check the LHT model.

\begin{table*}
\caption{\it The numbers of the signal and background events  after a series of kinematic cuts at the 14~TeV LHC with an integrated luminosity of 100 fb$^{-1}$. For the signal event $\kappa=3$. }
\label{tab:events}
\begin{tabular}{c|c|c|c|c|c|c|c} 
\hline
 &  $f=0.5~{\rm TeV}$ & $f=1.0~{\rm TeV}$ &  $f=1.5~{\rm TeV}$&  $t\bar{t}$ & $WW$ &$WZ$ & $ZZ$  \\ \hline
No cut & 88698 &2735.04&244.32&$5.5\times10^7$&$9.54\times10^6$&$4.36\times10^6$&$1.25\times10^6$ \\ \hline
$\met$, $W_{\rm jet}$-tagging and $M_{JJ}$ &295.33 & 38.51 & 5.51   & 13.31 & 4.77 & 6.43 & 1.17 \\ \hline

\end{tabular}
\end{table*}

~\\
\noindent{\bf Collider Simulation:}~%
The collider signature of interests to us is two $W$-jets with large missing momentum $\met$ arising from the two invisible $A_H$'s. 
The $\not{\!\!\rm E}_T$ plays the key role of triggering the signal events. In order to mimic the signal events, the SM background should consist of $W$ or $Z$ bosons. We consider SM backgrounds as follows: i) the pair production of $WW$, $WZ$ and $ZZ$ bosons; ii) $t\bar{t}$ productions; iii) the associated production of a $W$ boson and multiple jets (denoted by $W$+jets); iv) the associated production of a $Z$ boson and multiple jets ($Z$+jets). The $W$+jets, $Z$+jets and triple gauge boson productions are negligible after kinematics cuts. 
We generate the $W_H^+ W_H^-$ production and background events at the parton level using MadEvent~\cite{Alwall:2007st} at the 14 TeV LHC and pass events to Pythia~\cite{Sjostrand:2014zea} for showering and hadronization. We utilize Delphes~\cite{deFavereau:2013fsa} to simulate detector smearing effects in accord to a fairly standard Gaussian-type detector resolution given by $\delta E/E= \mathcal{A}/\sqrt{E/{\rm GeV}}\oplus \mathcal{B}$,
where $\mathcal{A}$ is a sampling term and $\mathcal{B}$ is a constant term.  For leptons we take $\mathcal{A}=5\%$ and $\mathcal{B}=0.55\%$, and for jets we take $\mathcal{A}=100\%$ 
and $\mathcal{B}=5\%$.
We also impose the lepton veto if the lepton has transverse momentum $p^{\ell}_T$ greater than $20$ GeV, rapidity $\left|\eta_{\ell}\right|\leq 2.5~$ and its overlap with jets  $\Delta R_{j\ell} = \sqrt{(\Delta \eta)^2 + (\Delta \phi)^2} \geq 0.4$. The missing transverse momentum ($\met$) is then defined to balance the total transverse momentum of visible objects.

\begin{figure}
\includegraphics[scale=0.55]{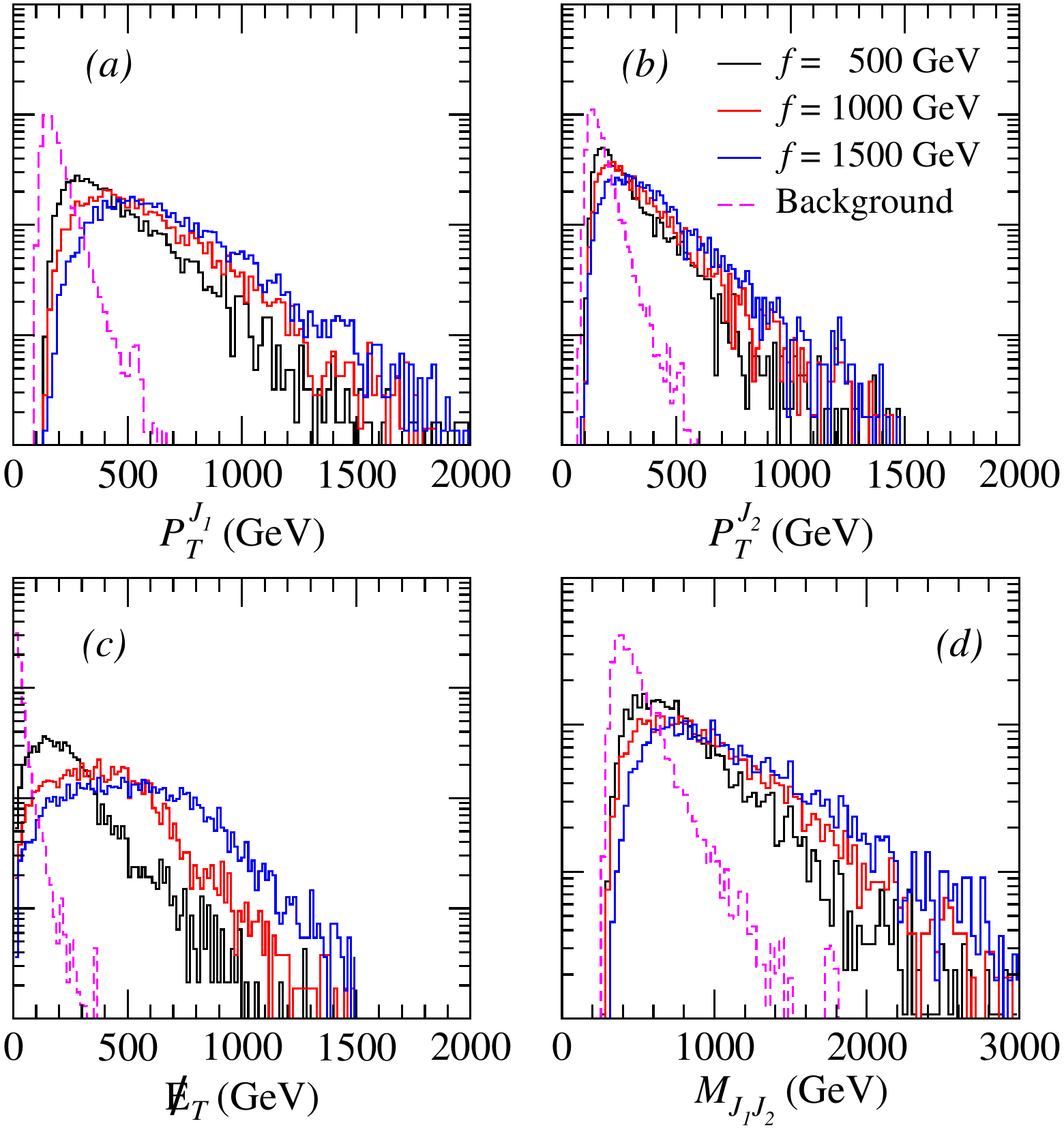}
\caption{Normalized distribution of the signal events ($f=0.5, 1.0, 1.5~{\rm TeV}$) and background events: $p_T$ of the leading $W$-jet (a) and the subleading $W$-jet (b); the invariant mass of the two $W$-jets (c); the missing transverse momentum (d). }
\label{fig:dist}
\end{figure}

A 2-pronged boosted $W$-jet is tagged using the so-called ``mass-drop" technique with asymmetry cut introduced in Ref.~\cite{Butterworth:2008iy}. The $W$-jet reconstruction is performed using Cambridge/Aachen algorithm with Fastjet~\cite{Cacciari:2011ma}. The distance parameter of $1.2$ is adopted to cluster a fat jet that is initiated by the boosted $W$ boson.  
We further require the invariant masses of those reconstructed fat 2-prong jets ($M_J$) within the following mass window~\cite{Aad:2015owa}:
\beq
m_W -13~{\rm GeV} \leq M_{J} \leq m_W + 13~{\rm GeV}, 
\eeq
where $m_W = 80.425~{\rm GeV}$ is the mass of $W$-boson. 
After $p_T$ ordering, we plot the $p_T$ distributions of the leading $W$-jet and the subleading $W$-jet of the signal and background events in Fig.~\ref{fig:dist}(a) and \ref{fig:dist}(b), respectively. Since the $W$-jets in the signal events arise from the heavy $W_H$ decays, their $p_T$ distributions are in general harder than those of the SM background events.
We demand the leading $W$-jet ($J_1$) and the subleading $W$-jet ($J_2$) to satisfy 
\beq
 p_T^{J_1}\geq 200~{\rm GeV}, ~p_T^{J_2}\geq 100~{\rm GeV},~\left| \eta^{J_{1,2}}\right| \leq 3,
\label{eqWtag:ptcut}
\eeq
to suppress the SM background. The numbers of events of the signal and background after the above cut are shown in the third row of Table~\ref{tab:events} with an integrated luminosity of $100~{\rm fb}^{-1}$. For illustration we choose three benchmark values of $f~(0.5~{\rm TeV},1~{\rm TeV},1.5~{\rm TeV})$. For simplification we fix $\kappa=3$ assuming the masses of T-odd fermions, $M_{f_-}\simeq \sqrt{2}\kappa f$, are much heavier than $W_H$. We emphasize that the cut efficiencies obtained above depend mainly on the mass split between $W_H$ and $A_H$. 

We take the advantage of the large missing transverse momentum of the signal events to trigger the signal events and reject the SM backgrounds; see Figure~\ref{fig:dist}(c). We  impose a hard cut on the missing transverse momentum as $\met \geq 400~{\rm GeV}$.
Furthermore, the two $W$-jets in the signal events originate from two $W_H$'s and exhibit a large invariant mass ($M_{JJ}$) as shown in Figure~\ref{fig:dist}(c). We thus impose a hard cut on $M_{J_1 J_2}$ as following: 
\beq
M_{J_1 J_2}\geq 500~{\rm GeV},
\eeq
which efficiently reduces the SM backgrounds; see the third row in Table~\ref{tab:events}. 

Given the cut efficiencies of both the signal and backgrounds, we are in a position to estimate the potential of the LHC in the search for $W_H^+W_H^-$ pair production with the signature of two fat $W$-jets and  large missing energy. We calculate the $5\sigma$ discovery potential using
\beq
\sqrt{-2\left[ (n_b+n_s) \ln \left(\frac{n_b}{n_s + n_b}\right) + n_s\right]}= 5
\eeq
where $n_s$ denotes the number of the signal events while $n_b$ the number of the background events~\cite{Cowan:2010js}.

\begin{figure}
\includegraphics[scale=0.44]{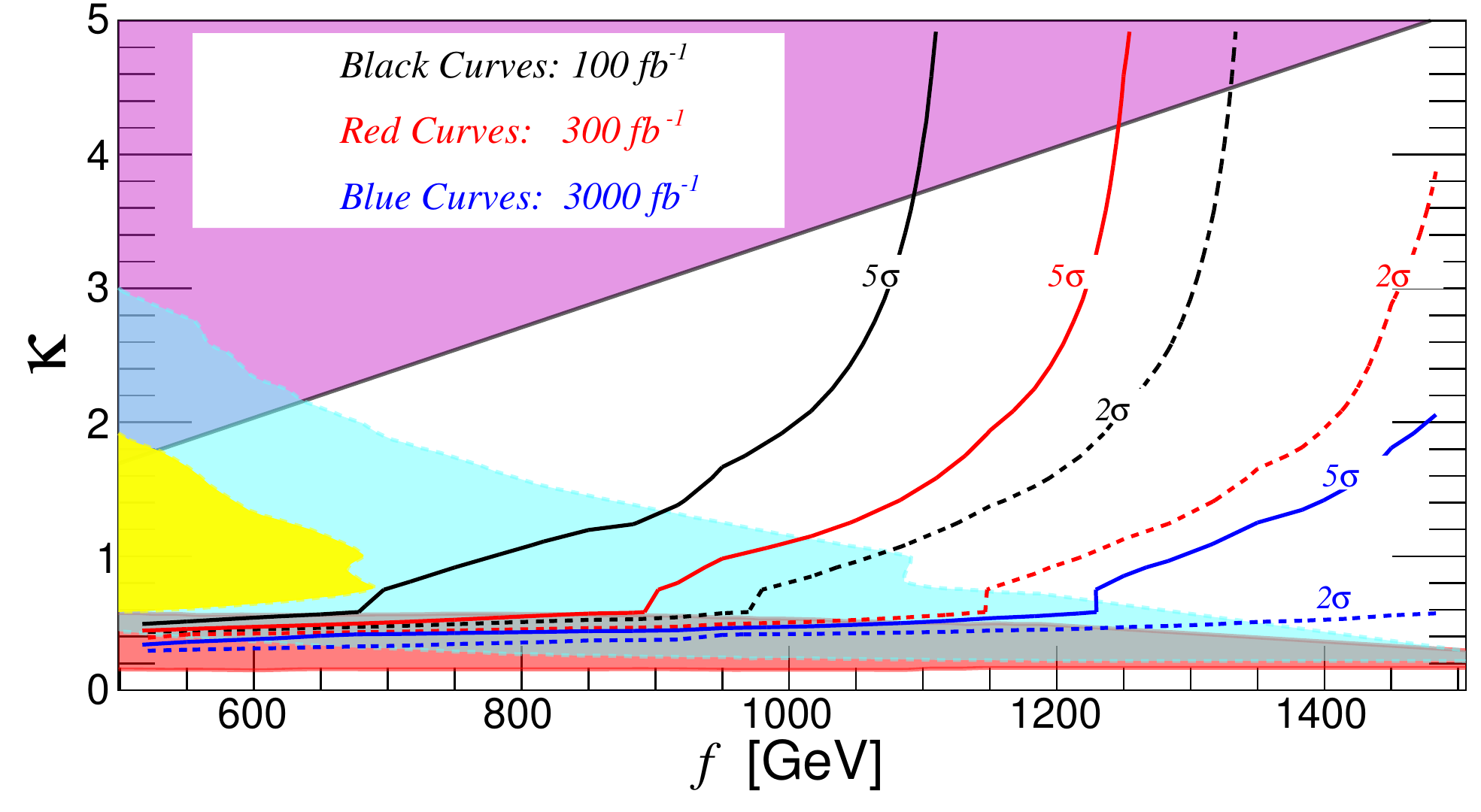}
\caption{The $5\sigma$ discovery potential (solid curves) and $2\sigma$ exclusion limit (dashed curves) of the $W_H^+W_H^-$ pair production at the 14~TeV LHC with three integrated luminosities: $\mathcal{L}=100~{\rm fb}^{-1}$ (black), $300~{\rm fb}^{-1}$ (red) and $3000~{\rm fb}^{-1}$ (blue). The shaded regions are excluded by the low energy experiments and the LHC Run-1 data~\cite{Reuter:2013iya}; see text for more details. 
} 
\label{fig:significance}
\end{figure}

The $5\sigma$ sensitivity in the plane of $f$ and $\kappa$ is shown in Fig.~\ref{fig:significance} for three benchmark integrated luminosities of  $\mathcal{L}=100~{\rm fb}^{-1}$, $300~{\rm fb}^{-1}$ and $3000~{\rm fb}^{-1}$.  For $\kappa\gtrsim 2.5$ the discovery potential is not sensitive to $f$.  It is owing to the fact that the $T$-odd fermion's contribution to the $W_H W_H$ pair production is highly suppressed for a large value of $\kappa$. As a result the production cross section depends mainly on the $f$ rather than $\kappa$. Obviously a large amount of parameter spaces could be probed at the high luminosity (HL, $3000~{\rm fb}^{-1}$) LHC. For comparison, we also plot the current experimental limits~\cite{Reuter:2013iya}: 
the magenta region denotes the indirect constraint from the four-fermion contact operators $eedd$; the blue region is constrained by the Jet and $\met$ data, the yellow region is bounded by the Leptons, Jets and $\met$ and the red region at the bottom is excluded by the Monojet and $\met$; see Ref.~\cite{Reuter:2013iya} for details. Most of the allowed parameter spaces can be probed in the $W_HW_H$ pair production at the high luminosity LHC with an integrated luminosity of $3000~{\rm fb}^{-1}$, see the blue solid line.

The $W_H^+W_H^-$ pair production is also a good process for excluding the parameter space of the LHT model.  Figure~\ref{fig:significance} also shows the $2\sigma$ exclusion limits in the plane of $f$ and $\kappa$, which are calculated using 
\beq
\sqrt{-2 \left(n_b \ln \frac{n_s + n_b}{n_b}-n_s \right)} = 1.96~. 
\eeq
The red (black, blue) dashed curve denotes the exclusion limit at the LHC with an integrated luminosity of $100~{\rm fb}^{-1}$ ($300~{\rm fb }^{-1}$, $3000~{\rm fb}^{-1}$). If, unfortunately, no excess were observed in the $W^+_H W^-_H$ pair production at the HL-LHC, then we conclude that almost entirely allowed parameter space of $f\leq 1.5~{\rm TeV}$ by the current data would be ruled out; see the blue dashed curve. A large $f$, say $f>1.5~{\rm TeV}$, would reintroduce Little Hierarchy problem which weaken the original motivation of the LHT model.

\begin{figure}[b]
\includegraphics[scale=0.5]{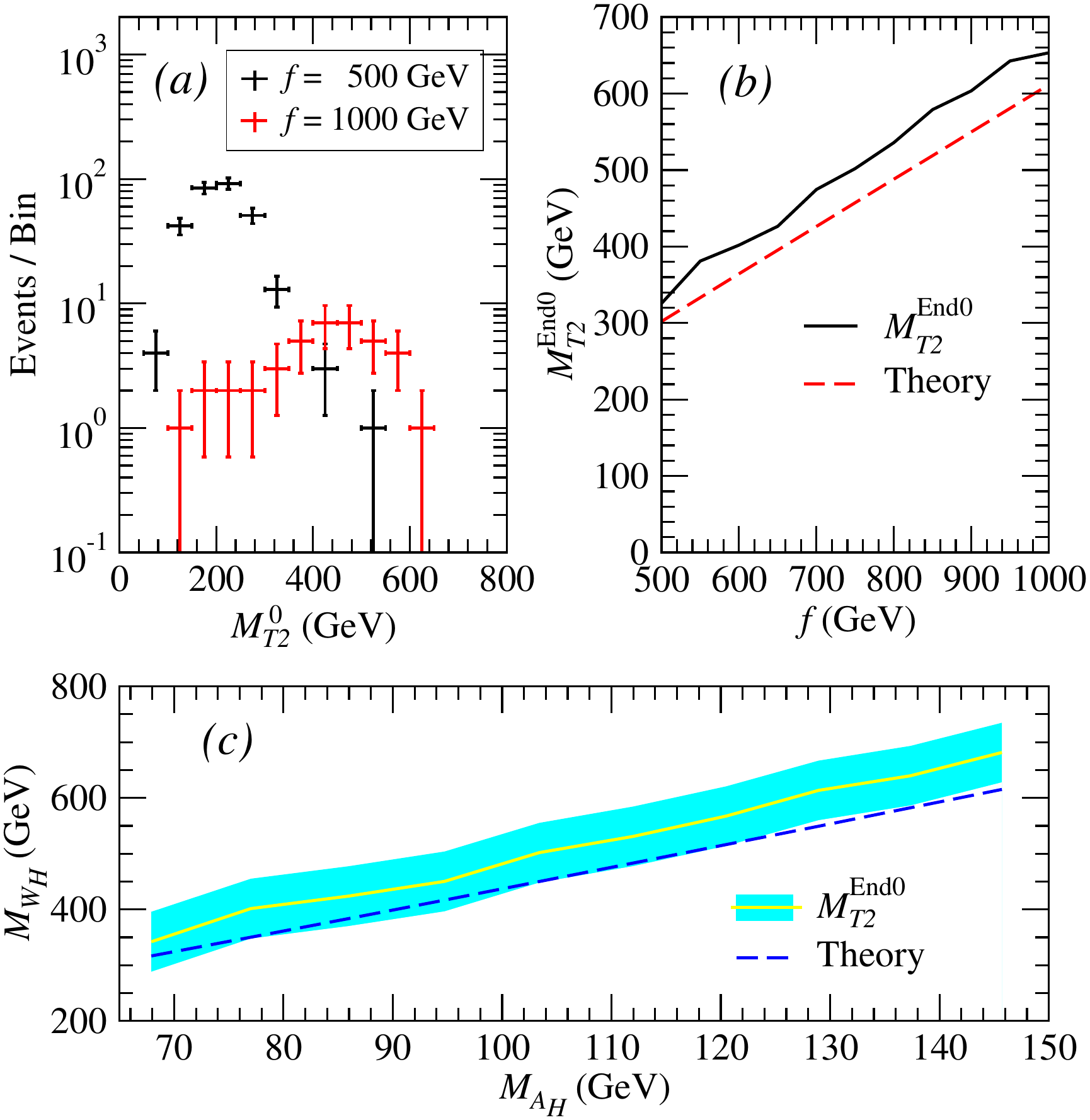}
\caption{(a) The $M^0_{T2}$ distribution for $f=0.5~(1.0)$ TeV for an integrated luminosity of $100~{\rm fb}^{-1}$; (b) the $M_{T2}^{\rm End0}$ determined in the so-called ``kink-to-bump'' method~\cite{Curtin:2011ng} versus the scale $f$ (solid curve) while the true theoretical value of $M_{T2}^{\rm End}$ is also plotted for comparison (dashed curve).} 
\label{fig:mt2}
\end{figure}

\noindent{\bf The measurements of $f$ and $M_{W_H}$:}~%
Next we demonstrate that the parameters $f$, $M_{W_H}$ and $M_{A_H}$ in the LHT model can be determined from the $M^0_{T2}$ distribution where $M^0_{T2}=m_{T2}(p_{J_1}, p_{J_2}, \met; M_{A_H}=0)$, where $p_{J_1}$ and $p_{J_2}$ are the momentum of two $W$-jets. Figure~\ref{fig:mt2}(a) displays the $M_{T2}^0$ distribution for $f=500$ GeV and $1000$ GeV, respectively. Only statistical uncertainty is included in the plot.

The value of $M_{T2}^{\rm End0}$, the endpoint of $M_{T2}^0$, is extracted out by  a ``kink-to-bump" method~\cite{Curtin:2011ng}.
 The method utilizes two straight lines to fit the $M_{T2}^0$ distribution around the edge to find the kink position.  A global fitting then gives rise to a likelihood function peaking around the true kink position.
We obtain the values of $M_{T2}^{\rm End0}$ as 325 (653)~GeV for $f=500~(1000)~{\rm GeV}$, respectively. The theoretical values predicted by the LHT model in Eq.~(\ref{eq:mt2lht0}) are $302$ GeV and $611$ GeV, respectively. The measured values are in good agreement with the LHT predictions. Figure~\ref{fig:mt2}(b) plots the value of $M_{T2}^{\rm End0}$ as a function of $f$. The solid curve represents the $M_{T2}^{\rm End0}$ obtained from  fitting while the dashed curve denotes the  theoretical relation. The $\sim 10\%$ discrepancy can be treated as the uncertainty of our method. In practice one could use $M_{A_H}$ derived from $M_{T2}^{\rm End0}$ to reconstruct the $M_{T2}$ distribution and then measure $M_{W_H}$ from its endpoint. 

The $M_{W_H}$ and $M_{A_H}$ can be calculated directly from $f$ once the value of $f$ is determined from $M_{T2}^{\rm End0}$ measurement. So far only the statistical uncertainty is considered in the calculation. The systematical uncertainties are expected to be much larger than the statistical uncertainties, however. The light blue band in the Figure \ref{fig:mt2}(c) shows the calculated $W_H$ and $A_H$ when a $\pm 50~{\rm GeV}$ uncertainty is taken into account in   $M_{T2}^{\rm End0}$ measurement, which is in good agreement with the unique mass relation between T-odd heavy gauge bosons in LHT model shown by the dashed line.

~\\
\noindent{\bf Summary and Discussion:~}%
Little Higgs Model with T-parity provides an explanation for the Higgs mass being at the weak scale and a suitable candidate for dark matter particle. With the aid of jet substructure, LHC has the capability of probing most of the parameter space with $W_H$ pair production, using the hadronic final state. We propose a strategy to determine the most important parameter, the energy scale $f$, using the $M_{T2}$ analysis.  The unique mass relation between the heavy gauge bosons $W_H$ and $A_H$ can be tested as well.

A pair of boosted $W$-jets is an important channel to search for new resonance at the LHC Run-II ~\cite{Aad:2015owa,Khachatryan:2014hpa,CMS-PAS-EXO-15-002}. However, when dark matter is involved in the decay products, resonances of diboson pairs will not be reconstructed and our approach can be easily incorporated. As the 2~TeV diboson anomaly found in Run-I data is not confirmed by the Run-II data~\cite{ATLAS-CONF-2015-073}, we recommend that the data should be revisited with missing energy analysis, as we have shown in this letter.

~\\
\noindent{\bf Acknowledgments:~}%
CRC would like to acknowledge the support of National Center for Theoretical Sciences (NCTS).
The work of QHC and YDL is supported in part by the National Science Foundation of China under Grand No. 11275009. The work of CRC is supported in part by the National Science Council of R.O.C. under Grants No.~NSC 102-2112-M-003-001-MY3.

\bibliographystyle{apsrev}
\bibliography{reference}

\begin{thebibliography}{30}
\expandafter\ifx\csname natexlab\endcsname\relax\def\natexlab#1{#1}\fi
\expandafter\ifx\csname bibnamefont\endcsname\relax
  \def\bibnamefont#1{#1}\fi
\expandafter\ifx\csname bibfnamefont\endcsname\relax
  \def\bibfnamefont#1{#1}\fi
\expandafter\ifx\csname citenamefont\endcsname\relax
  \def\citenamefont#1{#1}\fi
\expandafter\ifx\csname url\endcsname\relax
  \def\url#1{\texttt{#1}}\fi
\expandafter\ifx\csname urlprefix\endcsname\relax\def\urlprefix{URL }\fi
\providecommand{\bibinfo}[2]{#2}
\providecommand{\eprint}[2][]{\url{#2}}

\bibitem[{\citenamefont{Arkani-Hamed et~al.}(2001)\citenamefont{Arkani-Hamed,
  Cohen, and Georgi}}]{ArkaniHamed:2001nc}
\bibinfo{author}{\bibfnamefont{N.}~\bibnamefont{Arkani-Hamed}},
  \bibinfo{author}{\bibfnamefont{A.~G.} \bibnamefont{Cohen}}, \bibnamefont{and}
  \bibinfo{author}{\bibfnamefont{H.}~\bibnamefont{Georgi}},
  \bibinfo{journal}{Phys. Lett.} \textbf{\bibinfo{volume}{B513}},
  \bibinfo{pages}{232} (\bibinfo{year}{2001}), \eprint{hep-ph/0105239}.

\bibitem[{\citenamefont{Arkani-Hamed et~al.}(2002)\citenamefont{Arkani-Hamed,
  Cohen, Katz, and Nelson}}]{ArkaniHamed:2002qy}
\bibinfo{author}{\bibfnamefont{N.}~\bibnamefont{Arkani-Hamed}},
  \bibinfo{author}{\bibfnamefont{A.~G.} \bibnamefont{Cohen}},
  \bibinfo{author}{\bibfnamefont{E.}~\bibnamefont{Katz}}, \bibnamefont{and}
  \bibinfo{author}{\bibfnamefont{A.~E.} \bibnamefont{Nelson}},
  \bibinfo{journal}{JHEP} \textbf{\bibinfo{volume}{07}}, \bibinfo{pages}{034}
  (\bibinfo{year}{2002}), \eprint{hep-ph/0206021}.

\bibitem[{\citenamefont{Csaki et~al.}(2003)\citenamefont{Csaki, Hubisz, Kribs,
  Meade, and Terning}}]{Csaki:2002qg}
\bibinfo{author}{\bibfnamefont{C.}~\bibnamefont{Csaki}},
  \bibinfo{author}{\bibfnamefont{J.}~\bibnamefont{Hubisz}},
  \bibinfo{author}{\bibfnamefont{G.~D.} \bibnamefont{Kribs}},
  \bibinfo{author}{\bibfnamefont{P.}~\bibnamefont{Meade}}, \bibnamefont{and}
  \bibinfo{author}{\bibfnamefont{J.}~\bibnamefont{Terning}},
  \bibinfo{journal}{Phys. Rev.} \textbf{\bibinfo{volume}{D67}},
  \bibinfo{pages}{115002} (\bibinfo{year}{2003}), \eprint{hep-ph/0211124}.

\bibitem[{\citenamefont{Cheng and Low}(2003)}]{Cheng:2003ju}
\bibinfo{author}{\bibfnamefont{H.-C.} \bibnamefont{Cheng}} \bibnamefont{and}
  \bibinfo{author}{\bibfnamefont{I.}~\bibnamefont{Low}},
  \bibinfo{journal}{JHEP} \textbf{\bibinfo{volume}{09}}, \bibinfo{pages}{051}
  (\bibinfo{year}{2003}), \eprint{hep-ph/0308199}.

\bibitem[{\citenamefont{Cheng and Low}(2004)}]{Cheng:2004yc}
\bibinfo{author}{\bibfnamefont{H.-C.} \bibnamefont{Cheng}} \bibnamefont{and}
  \bibinfo{author}{\bibfnamefont{I.}~\bibnamefont{Low}},
  \bibinfo{journal}{JHEP} \textbf{\bibinfo{volume}{08}}, \bibinfo{pages}{061}
  (\bibinfo{year}{2004}), \eprint{hep-ph/0405243}.

\bibitem[{\citenamefont{Low}(2004)}]{Low:2004xc}
\bibinfo{author}{\bibfnamefont{I.}~\bibnamefont{Low}}, \bibinfo{journal}{JHEP}
  \textbf{\bibinfo{volume}{10}}, \bibinfo{pages}{067} (\bibinfo{year}{2004}),
  \eprint{hep-ph/0409025}.

\bibitem[{\citenamefont{Hubisz et~al.}(2006)\citenamefont{Hubisz, Meade, Noble,
  and Perelstein}}]{Hubisz:2005tx}
\bibinfo{author}{\bibfnamefont{J.}~\bibnamefont{Hubisz}},
  \bibinfo{author}{\bibfnamefont{P.}~\bibnamefont{Meade}},
  \bibinfo{author}{\bibfnamefont{A.}~\bibnamefont{Noble}}, \bibnamefont{and}
  \bibinfo{author}{\bibfnamefont{M.}~\bibnamefont{Perelstein}},
  \bibinfo{journal}{JHEP} \textbf{\bibinfo{volume}{01}}, \bibinfo{pages}{135}
  (\bibinfo{year}{2006}), \eprint{hep-ph/0506042}.

\bibitem[{\citenamefont{Hubisz and Meade}(2005)}]{Hubisz:2004ft}
\bibinfo{author}{\bibfnamefont{J.}~\bibnamefont{Hubisz}} \bibnamefont{and}
  \bibinfo{author}{\bibfnamefont{P.}~\bibnamefont{Meade}},
  \bibinfo{journal}{Phys. Rev.} \textbf{\bibinfo{volume}{D71}},
  \bibinfo{pages}{035016} (\bibinfo{year}{2005}), \eprint{hep-ph/0411264}.

\bibitem[{\citenamefont{Belyaev et~al.}(2006)\citenamefont{Belyaev, Chen, Tobe,
  and Yuan}}]{Belyaev:2006jh}
\bibinfo{author}{\bibfnamefont{A.}~\bibnamefont{Belyaev}},
  \bibinfo{author}{\bibfnamefont{C.-R.} \bibnamefont{Chen}},
  \bibinfo{author}{\bibfnamefont{K.}~\bibnamefont{Tobe}}, \bibnamefont{and}
  \bibinfo{author}{\bibfnamefont{C.~P.} \bibnamefont{Yuan}},
  \bibinfo{journal}{Phys. Rev.} \textbf{\bibinfo{volume}{D74}},
  \bibinfo{pages}{115020} (\bibinfo{year}{2006}), \eprint{hep-ph/0609179}.

\bibitem[{\citenamefont{Reuter et~al.}(2014)\citenamefont{Reuter, Tonini, and
  de~Vries}}]{Reuter:2013iya}
\bibinfo{author}{\bibfnamefont{J.}~\bibnamefont{Reuter}},
  \bibinfo{author}{\bibfnamefont{M.}~\bibnamefont{Tonini}}, \bibnamefont{and}
  \bibinfo{author}{\bibfnamefont{M.}~\bibnamefont{de~Vries}},
  \bibinfo{journal}{JHEP} \textbf{\bibinfo{volume}{02}}, \bibinfo{pages}{053}
  (\bibinfo{year}{2014}), \eprint{1310.2918}.

\bibitem[{\citenamefont{Asano et~al.}(2007)\citenamefont{Asano, Matsumoto,
  Okada, and Okada}}]{Asano:2006nr}
\bibinfo{author}{\bibfnamefont{M.}~\bibnamefont{Asano}},
  \bibinfo{author}{\bibfnamefont{S.}~\bibnamefont{Matsumoto}},
  \bibinfo{author}{\bibfnamefont{N.}~\bibnamefont{Okada}}, \bibnamefont{and}
  \bibinfo{author}{\bibfnamefont{Y.}~\bibnamefont{Okada}},
  \bibinfo{journal}{Phys. Rev.} \textbf{\bibinfo{volume}{D75}},
  \bibinfo{pages}{063506} (\bibinfo{year}{2007}), \eprint{hep-ph/0602157}.

\bibitem[{\citenamefont{Cao and Chen}(2007)}]{Cao:2007pv}
\bibinfo{author}{\bibfnamefont{Q.-H.} \bibnamefont{Cao}} \bibnamefont{and}
  \bibinfo{author}{\bibfnamefont{C.-R.} \bibnamefont{Chen}},
  \bibinfo{journal}{Phys. Rev.} \textbf{\bibinfo{volume}{D76}},
  \bibinfo{pages}{075007} (\bibinfo{year}{2007}), \eprint{0707.0877}.

\bibitem[{\citenamefont{Song-Ming et~al.}(2012)\citenamefont{Song-Ming, Lei,
  Wen, Wen-Gan, and Ren-You}}]{SongMing:2012gb}
\bibinfo{author}{\bibfnamefont{D.}~\bibnamefont{Song-Ming}},
  \bibinfo{author}{\bibfnamefont{G.}~\bibnamefont{Lei}},
  \bibinfo{author}{\bibfnamefont{L.}~\bibnamefont{Wen}},
  \bibinfo{author}{\bibfnamefont{M.}~\bibnamefont{Wen-Gan}}, \bibnamefont{and}
  \bibinfo{author}{\bibfnamefont{Z.}~\bibnamefont{Ren-You}},
  \bibinfo{journal}{Phys. Rev.} \textbf{\bibinfo{volume}{D86}},
  \bibinfo{pages}{054027} (\bibinfo{year}{2012}), \eprint{1208.5532}.

\bibitem[{\citenamefont{Khachatryan
  et~al.}(2014{\natexlab{a}})}]{Khachatryan:2014vla}
\bibinfo{author}{\bibfnamefont{V.}~\bibnamefont{Khachatryan}}
  \bibnamefont{et~al.} (\bibinfo{collaboration}{CMS}), \bibinfo{journal}{JHEP}
  \textbf{\bibinfo{volume}{12}}, \bibinfo{pages}{017}
  (\bibinfo{year}{2014}{\natexlab{a}}), \eprint{1410.4227}.

\bibitem[{\citenamefont{Aad et~al.}(2015)}]{Aad:2015owa}
\bibinfo{author}{\bibfnamefont{G.}~\bibnamefont{Aad}} \bibnamefont{et~al.}
  (\bibinfo{collaboration}{ATLAS}) (\bibinfo{year}{2015}), \eprint{1506.00962}.

\bibitem[{\citenamefont{Khachatryan
  et~al.}(2014{\natexlab{b}})}]{Khachatryan:2014hpa}
\bibinfo{author}{\bibfnamefont{V.}~\bibnamefont{Khachatryan}}
  \bibnamefont{et~al.} (\bibinfo{collaboration}{CMS}), \bibinfo{journal}{JHEP}
  \textbf{\bibinfo{volume}{08}}, \bibinfo{pages}{173}
  (\bibinfo{year}{2014}{\natexlab{b}}), \eprint{1405.1994}.

\bibitem[{\citenamefont{Collaboration}(2015)}]{CMS-PAS-EXO-15-002}
\bibinfo{author}{\bibfnamefont{C.}~\bibnamefont{Collaboration}}
  (\bibinfo{collaboration}{CMS}) (\bibinfo{year}{2015}),
  \eprint{CMS-PAS-EXO-15-002}.

\bibitem[{\citenamefont{Berger et~al.}(2010)\citenamefont{Berger, Cao, Chen,
  Shaughnessy, and Zhang}}]{Berger:2010fy}
\bibinfo{author}{\bibfnamefont{E.~L.} \bibnamefont{Berger}},
  \bibinfo{author}{\bibfnamefont{Q.-H.} \bibnamefont{Cao}},
  \bibinfo{author}{\bibfnamefont{C.-R.} \bibnamefont{Chen}},
  \bibinfo{author}{\bibfnamefont{G.}~\bibnamefont{Shaughnessy}},
  \bibnamefont{and} \bibinfo{author}{\bibfnamefont{H.}~\bibnamefont{Zhang}},
  \bibinfo{journal}{Phys. Rev. Lett.} \textbf{\bibinfo{volume}{105}},
  \bibinfo{pages}{181802} (\bibinfo{year}{2010}), \eprint{1005.2622}.

\bibitem[{\citenamefont{Zhang et~al.}(2011)\citenamefont{Zhang, Berger, Cao,
  Chen, and Shaughnessy}}]{Zhang:2010kr}
\bibinfo{author}{\bibfnamefont{H.}~\bibnamefont{Zhang}},
  \bibinfo{author}{\bibfnamefont{E.~L.} \bibnamefont{Berger}},
  \bibinfo{author}{\bibfnamefont{Q.-H.} \bibnamefont{Cao}},
  \bibinfo{author}{\bibfnamefont{C.-R.} \bibnamefont{Chen}}, \bibnamefont{and}
  \bibinfo{author}{\bibfnamefont{G.}~\bibnamefont{Shaughnessy}},
  \bibinfo{journal}{Phys. Lett.} \textbf{\bibinfo{volume}{B696}},
  \bibinfo{pages}{68} (\bibinfo{year}{2011}), \eprint{1009.5379}.

\bibitem[{\citenamefont{Cao et~al.}(2005)\citenamefont{Cao, Schwienhorst,
  Benitez, Brock, and Yuan}}]{Cao:2005pq}
\bibinfo{author}{\bibfnamefont{Q.-H.} \bibnamefont{Cao}},
  \bibinfo{author}{\bibfnamefont{R.}~\bibnamefont{Schwienhorst}},
  \bibinfo{author}{\bibfnamefont{J.~A.} \bibnamefont{Benitez}},
  \bibinfo{author}{\bibfnamefont{R.}~\bibnamefont{Brock}}, \bibnamefont{and}
  \bibinfo{author}{\bibfnamefont{C.~P.} \bibnamefont{Yuan}},
  \bibinfo{journal}{Phys. Rev.} \textbf{\bibinfo{volume}{D72}},
  \bibinfo{pages}{094027} (\bibinfo{year}{2005}), \eprint{hep-ph/0504230}.

\bibitem[{\citenamefont{Lester and Summers}(1999)}]{Lester:1999tx}
\bibinfo{author}{\bibfnamefont{C.~G.} \bibnamefont{Lester}} \bibnamefont{and}
  \bibinfo{author}{\bibfnamefont{D.~J.} \bibnamefont{Summers}},
  \bibinfo{journal}{Phys. Lett.} \textbf{\bibinfo{volume}{B463}},
  \bibinfo{pages}{99} (\bibinfo{year}{1999}), \eprint{hep-ph/9906349}.

\bibitem[{\citenamefont{Cho et~al.}(2008)\citenamefont{Cho, Choi, Kim, and
  Park}}]{Cho:2007dh}
\bibinfo{author}{\bibfnamefont{W.~S.} \bibnamefont{Cho}},
  \bibinfo{author}{\bibfnamefont{K.}~\bibnamefont{Choi}},
  \bibinfo{author}{\bibfnamefont{Y.~G.} \bibnamefont{Kim}}, \bibnamefont{and}
  \bibinfo{author}{\bibfnamefont{C.~B.} \bibnamefont{Park}},
  \bibinfo{journal}{JHEP} \textbf{\bibinfo{volume}{02}}, \bibinfo{pages}{035}
  (\bibinfo{year}{2008}), \eprint{0711.4526}.

\bibitem[{\citenamefont{Alwall et~al.}(2007)\citenamefont{Alwall, Demin,
  de~Visscher, Frederix, Herquet, Maltoni, Plehn, Rainwater, and
  Stelzer}}]{Alwall:2007st}
\bibinfo{author}{\bibfnamefont{J.}~\bibnamefont{Alwall}},
  \bibinfo{author}{\bibfnamefont{P.}~\bibnamefont{Demin}},
  \bibinfo{author}{\bibfnamefont{S.}~\bibnamefont{de~Visscher}},
  \bibinfo{author}{\bibfnamefont{R.}~\bibnamefont{Frederix}},
  \bibinfo{author}{\bibfnamefont{M.}~\bibnamefont{Herquet}},
  \bibinfo{author}{\bibfnamefont{F.}~\bibnamefont{Maltoni}},
  \bibinfo{author}{\bibfnamefont{T.}~\bibnamefont{Plehn}},
  \bibinfo{author}{\bibfnamefont{D.~L.} \bibnamefont{Rainwater}},
  \bibnamefont{and} \bibinfo{author}{\bibfnamefont{T.}~\bibnamefont{Stelzer}},
  \bibinfo{journal}{JHEP} \textbf{\bibinfo{volume}{09}}, \bibinfo{pages}{028}
  (\bibinfo{year}{2007}), \eprint{0706.2334}.

\bibitem[{\citenamefont{Sj{\"o}strand et~al.}(2015)\citenamefont{Sj{\"o}strand,
  Ask, Christiansen, Corke, Desai, Ilten, Mrenna, Prestel, Rasmussen, and
  Skands}}]{Sjostrand:2014zea}
\bibinfo{author}{\bibfnamefont{T.}~\bibnamefont{Sj{\"o}strand}},
  \bibinfo{author}{\bibfnamefont{S.}~\bibnamefont{Ask}},
  \bibinfo{author}{\bibfnamefont{J.~R.} \bibnamefont{Christiansen}},
  \bibinfo{author}{\bibfnamefont{R.}~\bibnamefont{Corke}},
  \bibinfo{author}{\bibfnamefont{N.}~\bibnamefont{Desai}},
  \bibinfo{author}{\bibfnamefont{P.}~\bibnamefont{Ilten}},
  \bibinfo{author}{\bibfnamefont{S.}~\bibnamefont{Mrenna}},
  \bibinfo{author}{\bibfnamefont{S.}~\bibnamefont{Prestel}},
  \bibinfo{author}{\bibfnamefont{C.~O.} \bibnamefont{Rasmussen}},
  \bibnamefont{and} \bibinfo{author}{\bibfnamefont{P.~Z.}
  \bibnamefont{Skands}}, \bibinfo{journal}{Comput. Phys. Commun.}
  \textbf{\bibinfo{volume}{191}}, \bibinfo{pages}{159} (\bibinfo{year}{2015}),
  \eprint{1410.3012}.

\bibitem[{\citenamefont{de~Favereau et~al.}(2014)\citenamefont{de~Favereau,
  Delaere, Demin, Giammanco, Lemaître, Mertens, and
  Selvaggi}}]{deFavereau:2013fsa}
\bibinfo{author}{\bibfnamefont{J.}~\bibnamefont{de~Favereau}},
  \bibinfo{author}{\bibfnamefont{C.}~\bibnamefont{Delaere}},
  \bibinfo{author}{\bibfnamefont{P.}~\bibnamefont{Demin}},
  \bibinfo{author}{\bibfnamefont{A.}~\bibnamefont{Giammanco}},
  \bibinfo{author}{\bibfnamefont{V.}~\bibnamefont{Lemaître}},
  \bibinfo{author}{\bibfnamefont{A.}~\bibnamefont{Mertens}}, \bibnamefont{and}
  \bibinfo{author}{\bibfnamefont{M.}~\bibnamefont{Selvaggi}}
  (\bibinfo{collaboration}{DELPHES 3}), \bibinfo{journal}{JHEP}
  \textbf{\bibinfo{volume}{02}}, \bibinfo{pages}{057} (\bibinfo{year}{2014}),
  \eprint{1307.6346}.

\bibitem[{\citenamefont{Butterworth et~al.}(2008)\citenamefont{Butterworth,
  Davison, Rubin, and Salam}}]{Butterworth:2008iy}
\bibinfo{author}{\bibfnamefont{J.~M.} \bibnamefont{Butterworth}},
  \bibinfo{author}{\bibfnamefont{A.~R.} \bibnamefont{Davison}},
  \bibinfo{author}{\bibfnamefont{M.}~\bibnamefont{Rubin}}, \bibnamefont{and}
  \bibinfo{author}{\bibfnamefont{G.~P.} \bibnamefont{Salam}},
  \bibinfo{journal}{Phys. Rev. Lett.} \textbf{\bibinfo{volume}{100}},
  \bibinfo{pages}{242001} (\bibinfo{year}{2008}), \eprint{0802.2470}.

\bibitem[{\citenamefont{Cacciari et~al.}(2012)\citenamefont{Cacciari, Salam,
  and Soyez}}]{Cacciari:2011ma}
\bibinfo{author}{\bibfnamefont{M.}~\bibnamefont{Cacciari}},
  \bibinfo{author}{\bibfnamefont{G.~P.} \bibnamefont{Salam}}, \bibnamefont{and}
  \bibinfo{author}{\bibfnamefont{G.}~\bibnamefont{Soyez}},
  \bibinfo{journal}{Eur. Phys. J.} \textbf{\bibinfo{volume}{C72}},
  \bibinfo{pages}{1896} (\bibinfo{year}{2012}), \eprint{1111.6097}.

\bibitem[{\citenamefont{Cowan et~al.}(2011)\citenamefont{Cowan, Cranmer, Gross,
  and Vitells}}]{Cowan:2010js}
\bibinfo{author}{\bibfnamefont{G.}~\bibnamefont{Cowan}},
  \bibinfo{author}{\bibfnamefont{K.}~\bibnamefont{Cranmer}},
  \bibinfo{author}{\bibfnamefont{E.}~\bibnamefont{Gross}}, \bibnamefont{and}
  \bibinfo{author}{\bibfnamefont{O.}~\bibnamefont{Vitells}},
  \bibinfo{journal}{Eur. Phys. J.} \textbf{\bibinfo{volume}{C71}},
  \bibinfo{pages}{1554} (\bibinfo{year}{2011}), \bibinfo{note}{[Erratum: Eur.
  Phys. J.C73,2501(2013)]}, \eprint{1007.1727}.

\bibitem[{\citenamefont{Curtin}(2012)}]{Curtin:2011ng}
\bibinfo{author}{\bibfnamefont{D.}~\bibnamefont{Curtin}},
  \bibinfo{journal}{Phys. Rev.} \textbf{\bibinfo{volume}{D85}},
  \bibinfo{pages}{075004} (\bibinfo{year}{2012}), \eprint{1112.1095}.

\bibitem[{\citenamefont{${\rm
  The~ATLAS~collaboration}$}(2015)}]{ATLAS-CONF-2015-073}
\bibinfo{author}{\bibnamefont{${\rm The~ATLAS~collaboration}$}}
  (\bibinfo{year}{2015}), \eprint{ATLAS-CONF-2015-073}.

\end{thebibliography}

\end{document}